\documentclass[conference]{IEEEtran}
\IEEEoverridecommandlockouts
\usepackage{cite}
\usepackage{amsmath,amssymb,amsfonts}
\usepackage{algorithmic}
\usepackage{graphicx}
\usepackage{color,soul}
\usepackage{textcomp}
\usepackage{xcolor}
\usepackage{siunitx}
\usepackage{textcomp}
\usepackage[multiple]{footmisc}
\usepackage{subcaption}
\usepackage{multirow}
\usepackage{float}
\usepackage{booktabs}
\usepackage{textcomp}
\usepackage{wrapfig}
\def\BibTeX{{\rm B\kern-.05em{\sc i\kern-.025em b}\kern-.08em
    T\kern-.1667em\lower.7ex\hbox{E}\kern-.125emX}}

     \usepackage[compact]{titlesec}
     \titlespacing{\section}{0pt}{0.5ex}{0.5ex}
     \titlespacing{\subsection}{0pt}{0.5ex}{0.5ex}
     \usepackage[compact]{titlesec}
     \titlespacing{\section}{0pt}{0.5ex}{0.5ex}
     \titlespacing{\subsection}{0pt}{0.5ex}{0.5ex}
\begin{document}

\title{VCO-CARE: VCO-based Calibration-free Analog Readout for Electrodermal activity sensing

\thanks{This work has been supported by the Madrid Government (Comunidad de Madrid-Spain) under the Multiannual Agreement with UC3M (“Fostering Young Doctors Research”, ENDOMEDEA CM-UC3M), by the the Swiss NSF Edge-Companions project (GA No. 10002812), and in part by the Swiss State Secretariat for Education, Research, and Innovation (SERI) through the SwissChips Research Project, by the ETSII of UPM under the project ETSII-UPM25-PU01, by the CEI Grants Program of UPM, and by the Spanish Ministry of Science, Innovation and Universities through the project “Co-HIPSTER-OE”, PID2024-158251NA-C33. This research was partially conducted by ACCESS – AI Chip Center for Emerging Smart Systems, supported by the InnoHK initiative of the Innovation and Technology Commission of the Hong Kong Special Administrative Region Government.}
}

\author{
\IEEEauthorblockN{Leidy Mabel Alvero-Gonzalez\IEEEauthorrefmark{1}, 
Matias Miguez\IEEEauthorrefmark{2}, 
Eric Gutierrez\IEEEauthorrefmark{3}, Juan Sapriza\IEEEauthorrefmark{4},
Susana Patón \IEEEauthorrefmark{3}, \\David Atienza\IEEEauthorrefmark{4} and 
José Miranda\IEEEauthorrefmark{5}
}
\IEEEauthorblockA{
Email: \IEEEauthorrefmark{1}leidy.alverogonzalez@ams-osram.com, \IEEEauthorrefmark{2}mmiguez@ucu.edu.uy, \\
\IEEEauthorrefmark{3}{egutier,spaton}@ing.uc3m.es,
\IEEEauthorrefmark{4}{juan.sapriza,david.atienza}@epfl.ch,
\IEEEauthorrefmark{5}jose.miranda@upm.es
\\
\IEEEauthorrefmark
{1}\textit{Medical Imaging Analog Design, ams OSRAM, Spain}, \IEEEauthorrefmark{2}\textit{Universidad Católica del Uruguay, Uruguay},\\ 
\IEEEauthorrefmark{3}\textit{Electronics Technology Department, UC3M, Spain}, \IEEEauthorrefmark{4}\textit{Embedded Systems Laboratory (ESL), EPFL, Switzerland},\\ 
\IEEEauthorrefmark{5}\textit{Center of Industrial Electronics (CEI), UPM, Spain}
}}

\maketitle

\begin{abstract}
    
Continuous monitoring of electrodermal activity (EDA) through wearable devices has attracted much attention in recent times. However, the persistent challenge demands analog front-end (AFE) systems with high sensitivity, low power consumption, and minimal calibration requirements to ensure practical usability in wearable technologies. In response to this challenge, this research introduces VCO-CARE, a Voltage-Controlled Oscillator-based Analog Readout tailored for continuous EDA sensing. 
The results show that our system achieves an exceptional average sensitivity of up to \makebox{40 pS} within a 0–20 \SI{}{\micro\siemens} range and a negligible relative error of less than 0.0025\% for fixed-resistance. Furthermore, the proposed system consumes only an average of 2.3 \SI{}{\micro\watt} based on post-layout validations and introduces a low noise contribution, measuring only 0.8~\SI{}{\micro\volt}$_\textrm{rms}$ across the \makebox{0-1.5 Hz} EDA signal band. This research aims to drive the evolution of wearable sensors characterized by seamless adaptability to diverse users, minimal power consumption, and outstanding noise resilience.
\end{abstract}


\section{Introduction}

\IEEEPARstart{E}lectrodermal Activity (EDA), or Galvanic Skin Response (GSR), is a key physiological signal used in emotion recognition, stress detection, and clinical monitoring. Its continuous acquisition is essential for wearable and personalized healthcare systems~\cite{nagai2019galvanic,wen2014emotion,das2016emotion,dutta2022analysis,miranda2022bindi,kappeler2013towards}, but requires high sensitivity, low power, and adaptability without extensive calibration~\cite{calero2023self}. Conventional biomedical sensing pipelines amplify, filter, and digitize low-swing analog signals, typically using the Successive Approximation Register (SAR) or Sigma-Delta ADCs. Although effective, these architectures require complex circuitry, precise references, or a high level of oversampling~\cite{MITPoh,rajeshPPGheartRate}. For EDA specifically, many recent systems rely on runtime calibration techniques, such as gain switching or adaptive impedance control, to handle inter- and intra-subject variability~\cite{MITPoh,xing2018wearable,kim2020fully,banganho2021designAdaptive}. Although such strategies improve dynamic range, they increase power, complexity, and computational burden, limiting their practicality for seamless, continuous monitoring~\cite{calero2023self}.

In the context of this work, instead of processing the input signal in the voltage domain, we propose directly converting the input voltage into the frequency domain using a VCO. In fact, the high sensitivity of ring oscillators is of special interest for biomedical applications because the instrumentation amplifier connected to the sensor can be removed \cite{ericVCOinst}. Additionally, the intrinsic first-order sinc function of the VCO-based ADC \cite{gielenLP} enables filtering operation, resulting in a sensing architecture limited to the sensor, the VCO-based ADC, and the digital processing stage. 

Although VCO-based ADCs are well-suited for low-power biosignal acquisition, they face challenges related to phase noise and non-linearity. In low-bandwidth applications like EDA, flicker noise must be minimized through careful transistor sizing. Despite inherent non-linearity, the monotonic voltage-to-frequency response, determined by the front-end topology, preserves amplitude-based features, and the small voltage swings typical of EDA signals further reduce distortion. 
On this basis, this paper introduces VCO-CARE, a subject-independent VCO-based ADC sensor for continuous EDA monitoring. Our proposed AFE eliminates the need for runtime recalibration, ensuring seamless adaptation across operating conductance ranges. 
The main contributions of the paper are as follows:
\begin{itemize}
    \item Design of a novel, subject-independent, ultra-low power, and ultra-low noise exosomatic EDA AFE.
    \item Simulations with a relative error of 0.0025\% and an average sensitivity of 40 pS, with 131 pS as the worst case, throughout a 0–20 \SI{}{\micro\siemens} operating conductance range without the need for manual or runtime calibration. 
    \item Validation through simulations at the schematic and post-layout level and realistic physiological data sets, demonstrating the practical viability and robustness of the proposed approach.
\end{itemize}

The rest of the document is structured as follows. The proposed AFE, simulation, and post-layout validations are detailed in Section \ref{system}. The results with real EDA data are presented in Section~\ref{results}. Section~\ref{discussion} discusses and compares the performance with \makebox{the state-of-the-art}. Finally, {Section} \ref{conclusions} states the conclusions and future work. 

\section{Proposed system}
\label{system}

\begin{figure}[htbp]
    \centering
    \includegraphics[width=0.98\columnwidth]{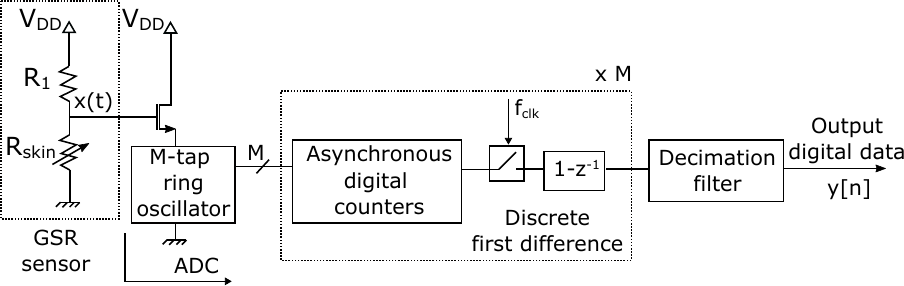}
    \caption{ Outline of VCO-CARE data acquisition AFE. 
    }
    \label{fig:proposed_system}
\end{figure}

The architecture of the proposed VCO-CARE acquisition system is illustrated in Fig.~\ref{fig:proposed_system}, which details the signal path from the GSR sensor to the digital output. Specifically, it begins with a resistive divider, where $R_\textrm{skin}$ represents the variable resistance of the skin, and $R_1$ is a fixed resistor selected to limit the current through the body. The voltage at the output of the divider, $x(t)$, is applied to the gate of a source follower-based ring oscillator. The oscillator converts \( x(t) \) into a frequency-modulated signal defined by \( f_\textrm{osc}(t) = f_\textrm{o} + K_\textrm{VCO} \cdot x(t) \), where \( f_\textrm{o} \) is the baseline frequency and \( K_\textrm{VCO} \) is the gain in Hz/V.

For the last stages, the VCO output is sampled by asynchronous digital counters at a clock rate $f_\textrm{clk}$, and then a discrete-time first difference is applied to extract the dynamic variations in frequency. Finally, a decimation filter reduces the output data rate, resulting in the final digital output $y[n]$.  

\subsection{VCO Design, Safety and Sensor Linearity Considerations}

\begin{figure}[htbp]
    \centering
    \includegraphics[width=0.85\columnwidth]{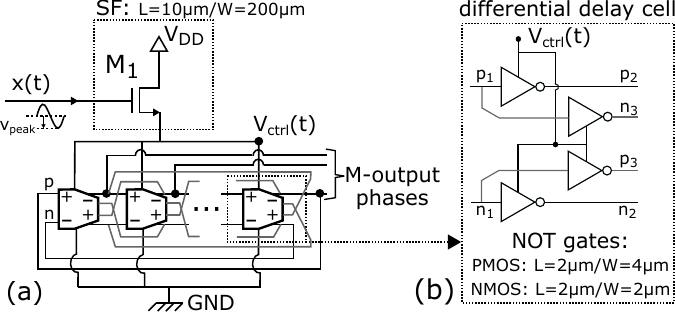}
    \caption{(a) Voltage-controlled oscillator implementation based on SF, (b) Differential delay cell for ring oscillators.}
    \label{fig:VCOdiffcells}
\end{figure}

The core of the proposed AFE is a 31-phase ring oscillator with pseudodifferential topology. The design follows the architecture described in \cite{borgmansFeedforwardRO} and is shown in Fig.\ref{fig:VCOdiffcells}(a). To boost the oscillation frequency,
the delay cell (Fig.~\ref{fig:VCOdiffcells}(b)) employs feedforward connections across multiple stages. This structure enables each phase to depend not only on its immediate predecessor but also on earlier phases, which improves the spectral properties. 
Note that for this work, the size of the device and the number of VCO taps were selected to balance two key aspects: low noise performance and manageable digital complexity. 

\begin{figure}[htbp]
    \centering
    \includegraphics[width=0.9\columnwidth]{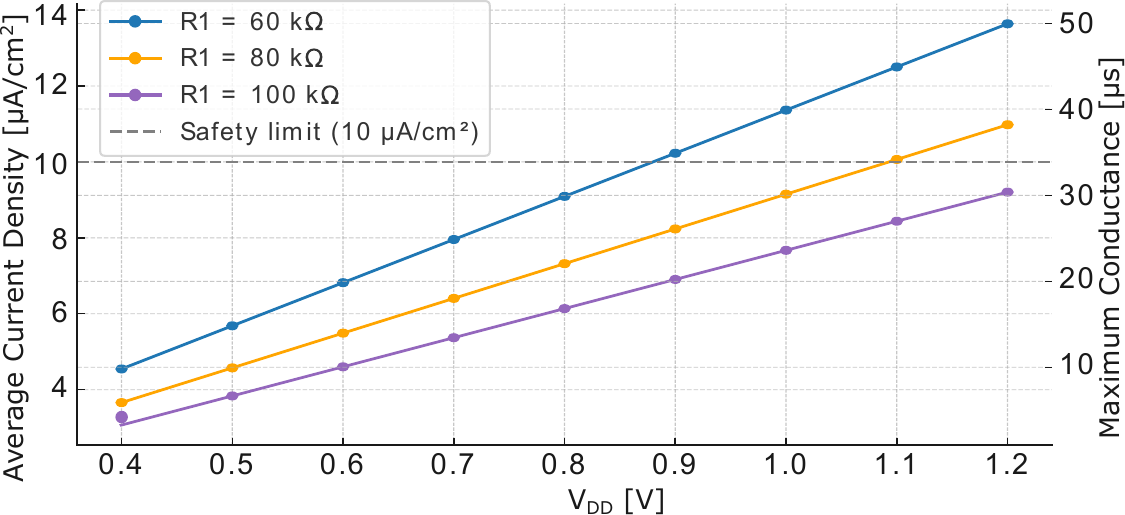}
    \caption{Average current density (left axis) and maximum measurable conductance before estimated VCO saturation (right axis) as a function of $V_\mathrm{DD}$ for different values of $R_1$, assuming a 1~cm$^2$ electrode area. 
    }
    \label{fig:range_current_tradeoff}
\end{figure}

\begin{figure}[htbp]
    \centering
    \includegraphics[width=0.9\columnwidth]{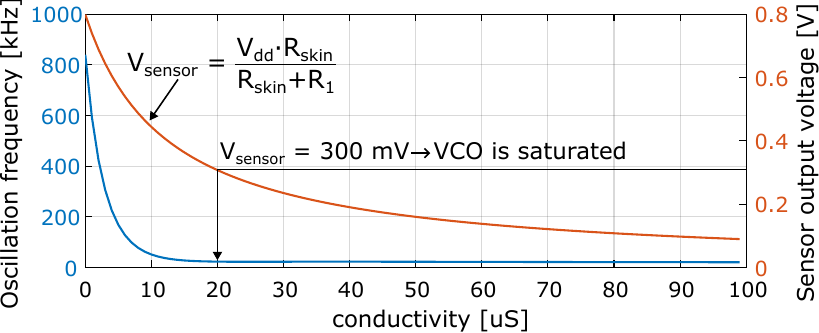}
    \caption{Voltage from the sensor vs. conductivity (orange) and oscillation frequency vs. conductivity (blue), for $R_1 = 80$~k$\Omega$. 
    }
    \label{fig:sensor_vcoFreq}
\end{figure}

Fig.~\ref{fig:range_current_tradeoff} illustrates how the design parameters $R_1$ and $V_\mathrm{DD}$ impact the trade-off between safety and sensing range. Increasing $V_\mathrm{DD}$ or decreasing $R_1$ extends the measurable conductance range, but also increases the current density, which may exceed the 10~\SI{}{\micro\ampere}/cm$^2$ safety limit. For this work, the selected configuration ($R_1 = 80$~k$\Omega$, $V_\mathrm{DD} = 0.8$~V) offers a practical compromise, enabling a conductance range of up to 20~\SI{}{\micro\siemens} while maintaining an average current density of 7.3~\SI{}{\micro\ampere}/cm$^2$. Note that this trade-off stems from the non-linear behavior of the sensor output voltage $x(t)$, derived from the resistive divider formed by $R_1$ and the skin resistance $R_\textrm{skin}$, as shown in Fig.~\ref{fig:sensor_vcoFreq}. Specifically, in our case, the ring oscillator saturates for input voltages below approximately 300~mV, flattening the voltage-to-frequency curve and degrading sensitivity. Although this threshold depends on the biasing of the front end and $V_\mathrm{DD}$, increasing the supply voltage to delay saturation would also increase power consumption and require recharacterization of the system.

\begin{figure}[htbp]
    \centering
    \includegraphics[width=0.95\columnwidth]{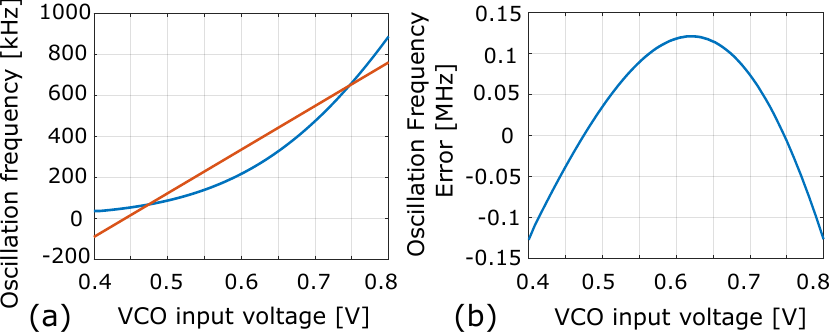}
    \caption{Static VCO non-linearity: (a) Voltage-to-frequency relation (blue) and linear fitting (orange); (b) Oscillation frequency error.}
    \label{fig:vco_polynomial}
\end{figure}

To further characterize the behavior of the VCO within the valid operating range, the plots in Fig.~\ref{fig:vco_polynomial} illustrate the static linearity in response to input voltage variations. This analysis was carried out on 
post-layout simulations using a typical transistor process and \SI{20}{\degree}C deg temperature, by sweeping the input voltage from 0~V to 0.8~V in 50~mV steps and extracting the corresponding oscillation frequency. The actual voltage-to-frequency transfer curve and its linear fit are shown in Fig.~\ref{fig:vco_polynomial}(a) for a wide input swing of 400~mV$_\textrm{pp}$. Fig.~\ref{fig:vco_polynomial}(b) quantifies the non-linearity through the deviation between the real and ideal responses. The VCO exhibits a maximum deviation of 130~kHz over the 36~kHz to 880~kHz range, corresponding to 16\% of the total oscillation range. Note that this is acceptable given the low bandwidth of the EDA signals and the monotonic nature of the VCO response, which preserves the integrity of the amplitude-based characteristics even in the presence of harmonic distortion. Finally, the extracted oscillation parameters are \makebox{f$_\textrm{o}$ = 220~kHz} and $K_\textrm{VCO}$ = 2.1~MHz/V.

Finally, the digital part is modeled and verified in QuestaSim\textregistered~ with a structure as proposed in \cite{quintero}. Specifically, we employ one of the phases of the ring oscillator to perform a coarse quantization of the output phase, and the rest of the phases to finely adjust the digital data. The selected sampling frequency is 12 Hz. The decimation stage corresponds to a moving average filter with a downsampling ratio of 4 to get the final digital data in a signal band of 1.5 Hz.

\subsection{Simulation results}
\label{sim_res}
Simulations are performed to validate the proper operation of the specific VCO-CARE implementation presented in this work. Due to the long time required for transient simulations in 65-nm, a model in MATLAB/Simulink\textregistered~ is used instead, adding the VCO non-idealities extracted from Cadence\textregistered Virtuoso\textregistered~ to the behavioral model: phase noise and nonlinear response. Nonlinearity is performed using a 5$^{th}$ degree polynomial approximation of the voltage-to-frequency conversion curve (Fig.~\ref{fig:vco_polynomial}(a) in blue). The VCO noise is generated considering the thermal noise and flicker noise components, which are 77~nV$_\textrm{rms}$ and 0.8~\SI{}{\micro\volt}$_\textrm{rms}$, respectively, over 1.5 Hz.

\begin{figure}[htbp]
    \centering
    \includegraphics[width=0.92\columnwidth]{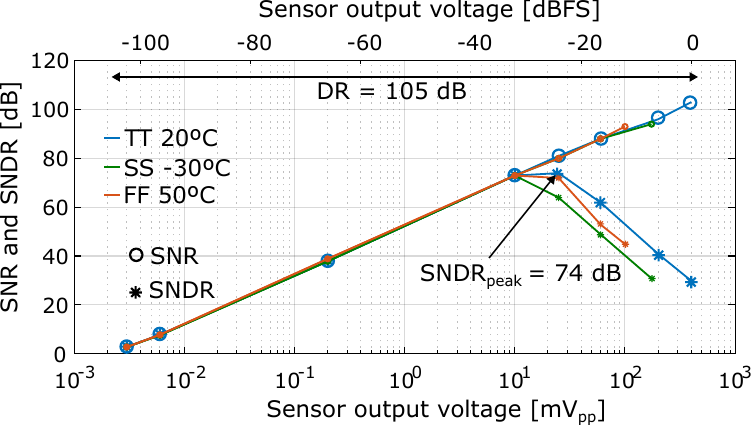}
    \caption{SNR, SNDR and DR plots of the pseudo-differential VCO-based ADC, for a single tone sinusoidal signal centered in 600~mV of the maximum swing of 400~mV$_\textrm{pp}$, at 0.33~Hz with V$_\textrm{DD}$ equal to 0.8~V.}
    \label{fig:ADC_performance}
\end{figure}

The dynamic performance depicted in Fig.~\ref{fig:ADC_performance} is evaluated by FFT computing under nominal conditions and the worst-case process corners (FF, SS) to obtain the output spectrum and analyze the noise and distortion. For typical process (TT) \SI{20}{\degree}Cdeg, the data converter achieves a dynamic range (DR) of 105~dB for 400~mV$_\textrm{pp}$ centered in 600~mV as the input signal, $x(t)$ in Fig.\ref{fig:VCOdiffcells}, which corresponds to 10~\SI{}{\micro\siemens} of conductivity. For voltage values larger than 60~mV$_\textrm{pp}$ the VCO becomes increasingly non-linear. The peak SNDR is then limited to 74~dB using 25~mV$_\textrm{pp}$ of voltage swing. Within this range of voltage, the VCO provides the best linearity behaviour. For the worst-cases corners, ADC's performance is not degraded dramatically. The noise floor is dominated by phase noise over quantization noise, with a total value of 0.805~\SI{}{\micro\volt}$_\textrm{rms}$ over bandwidth.

\begin{figure}[htbp]
    \centering
    \includegraphics[width=0.95\columnwidth]{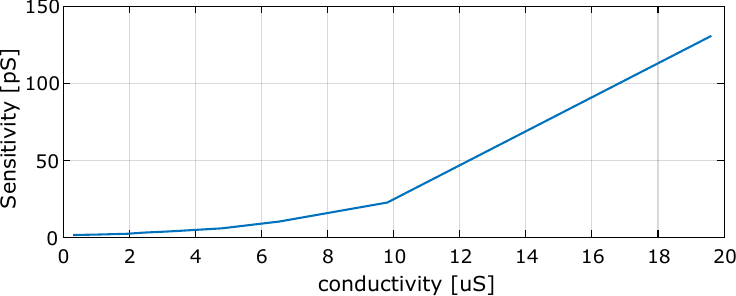}
    \caption{ADC sensitivity.}
    \label{fig:ADC_sensitivity}
\end{figure}

The ADC is statically characterised using several resistance values. 
Each of these corresponds directly to a voltage value obtained from the transfer function of the resistive divider as the input signal of the data converter. The measured resistance values are 
50, 101, 152, 208, 309, 409, 510, 1019, 2024, 2396, 3028, 3330, and 4031 k$\si{\ohm}$, i.e., approximately 0-20~\SI{}{\micro\siemens} range. 
The relative error of the converter data versus the measured conductivity is computed by removing the transfer function of the data converter in every code obtained for the resistances. The error values are extremely low, below 0.0025\%. Fig.~\ref{fig:ADC_sensitivity} shows the sensitivity in each measured conductance, defined by the slope of the converter, $\Delta$code/$\Delta$conductance. The average sensitivity is 40~pS, with 131~pS as the worst sensitivity obtained for 50~k$\Omega$, approximately 20~\SI{}{\micro\siemens}.

\subsection{Post-layout validation}
\label{layout_res}

\begin{figure}[htbp]
    \centering
    \includegraphics[width=0.98\columnwidth]{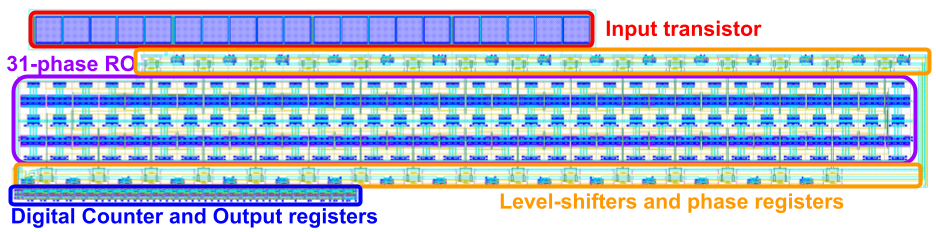}
    \caption{VCO custom layout in 65\SI{}{\nano\meter}, total area is 75~$\times$~340~\SI{}{\micro\meter}$^2$. Digital core
    not shown.
    }
    \label{fig:VCO_layout}
\end{figure}

The complete VCO subsystem, including the source follower input stage, the 31-phase ring oscillator, level shifters, the sense amplifiers, and the asynchronous counter with its corresponding registers that interface with the digital core, was implemented in 65~\SI{}{\nano\meter} CMOS technology. The VCO layout implementation preserves the oscillation parameters obtained at the schematic level. As shown in Fig.~\ref{fig:VCO_layout}, the layout occupies an area of 75~$\times$~340~\SI{}{\micro\meter}$^2$. 
The average value of power consumption is 1.2~\SI{}{\micro\watt}. 
The digital core was synthesized using Synopsys Design Compiler\textregistered, which produced an estimated power of 1.1~\SI{}{\micro\watt} at 1~V. Although the final power may increase in a tape-out, the combined estimated average consumption of 2.3~\SI{}{\micro\watt} remains highly competitive compared to the state of the art.

\section{Application example}
\label{results}

\begin{figure}[htbp]
    \centering
    \includegraphics[width=0.95\columnwidth]{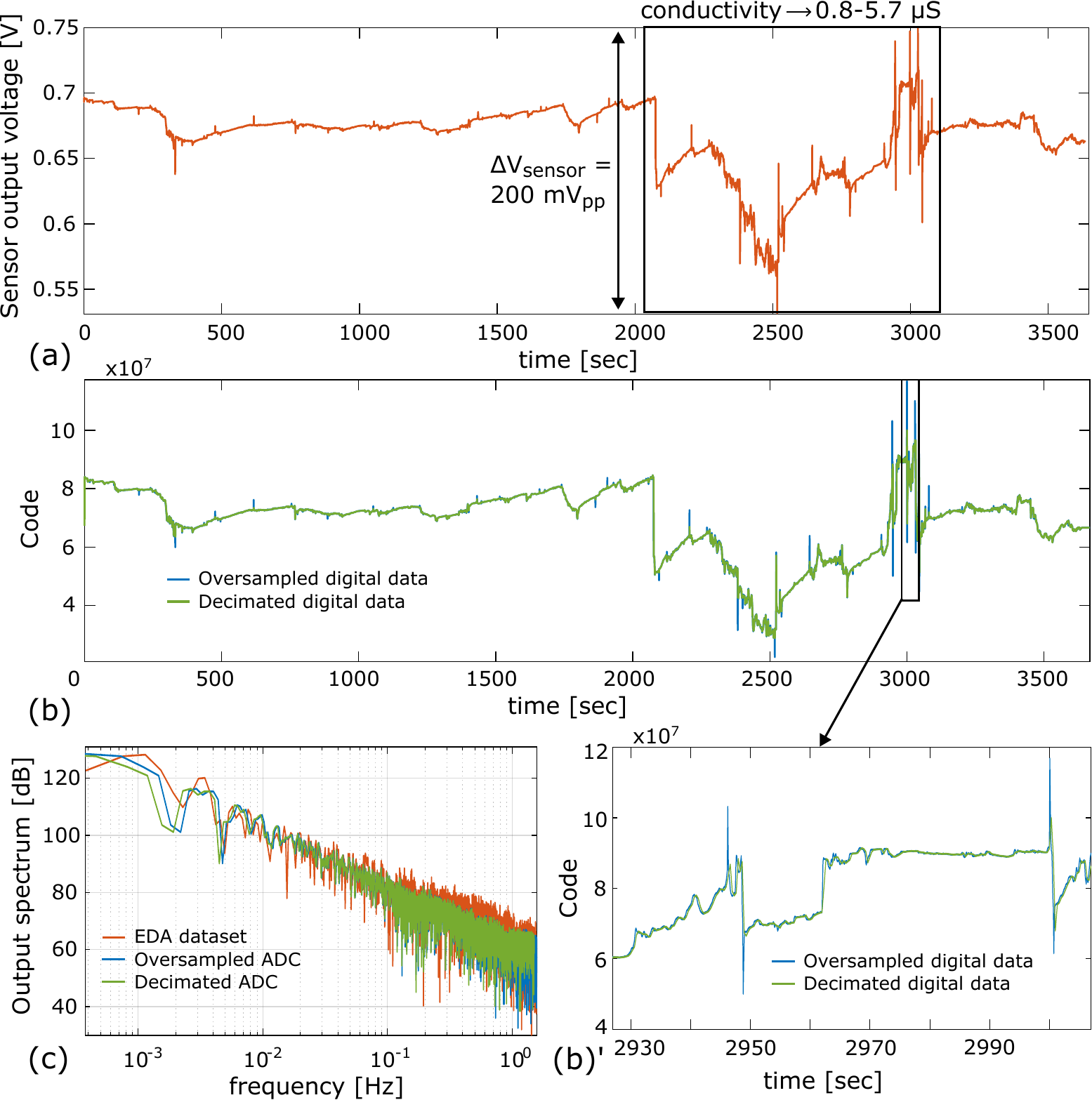}
    \caption{(a) EDA signal from the subject under several stimuli; (b) Oversampled (green) and decimated (green) digital data; (b)' Zoomed view of the digital data; (c) Output Spectra.}
    \label{fig:EDA_real_dataset}
\end{figure}

For a complete characterization of VCO-CARE, Fig. \ref{fig:EDA_real_dataset} shows the results using a dataset from a volunteer woman as input to the system. Note that the volunteer data were taken from \cite{miranda2022wemac} and introduced to our system (model with non-idealities used in \ref{sim_res}). The subject response was recorded evoking emotions using different stimuli (video clips) to identify variations of the EDA range under different conditions: fear vs. no fear. These situations can be distinguished by the EDA signal in orange in Fig. \ref{fig:EDA_real_dataset}(a). The first part of the signal shows a baseline condition; in contrast, in the last part, a stress response is observed, producing a wide variation in the sensor voltage of 200 mV$_\textrm{pp}$. The blue and green signals illustrated in Fig. \ref{fig:EDA_real_dataset}(b), corresponding to the oversampled and decimated digital output data, accurately follow the EDA signal. The zoomed view of the digital data in Fig. \ref{fig:EDA_real_dataset}(b)' depicts the smoothing function of the decimation filter, removing fluctuations from the oversampled data and retaining the EDA band. It is clearly appreciable in Fig. \ref{fig:EDA_real_dataset}(c) that the spectral information remains throughout the proposed system, with the spectra of the digitized signals and the EDA signal being identical. 

\section{Comparison to prior work}
\label{discussion}
Table~\ref{comparison} presents a comparative analysis between the proposed system and recent state-of-the-art EDA acquisition approaches~\cite{MITPoh,xing2018wearable,calero2023self}. Unlike these systems, which require runtime adjustments for gain or range calibration, the proposed design achieves subject-independence without any dynamic tuning within the operational conductance range. 

The proposed system achieves a mean sensitivity of 40~pS and a relative error below 0.0025\% within the 0–20~\SI{}{\micro\siemens} range. 
In contrast, \cite{calero2023self} reports a mean sensitivity of 130~pS (over the same range) and a relative error of approximately 1\%, highlighting a more than 8$\times$ improvement in sensitivity and two orders of magnitude lower error achieved by VCO-CARE. Moreover, while \cite{calero2023self} reports a power consumption of 620~\SI{}{\micro\watt}, the proposed approach is estimated to operate at just 2.3~\SI{}{\micro\watt}, marking a substantial reduction in energy demand. Other systems such as \cite{MITPoh,xing2018wearable} do not report complete power or error metrics or operate over narrower conductance ranges.

In general, the proposed architecture provides a competitive trade-off between sensitivity, accuracy, and power, while uniquely eliminating the need for calibration circuitry.

\begin{table}[htbp]
    \centering
    \caption{{Comparison to self-adjustable based EDA sensors.}}
    \begin{tabular}{|c|c|c|c|c|}
  \hline
	Work & \cite{MITPoh} & \cite{xing2018wearable} & \cite{calero2023self} & \textbf{This}\\
     \hline
	 Measurement & {Automatic}  & {Wheastone} & {Wheastone} & {VCO-based}\\
    technique & {bias} & {bridge} & {bridge} & {ADC} \\
         \hline
         Conductivity & \multirow{2}{*}{0.25-10} & \multirow{2}{*}{0.5-6} &  \multirow{2}{*}{0-40} & \multirow{2}{*}{0-20} \\
     Range {[\SI{}{\micro\siemens}]} & {} & {} & {} & {}\\ 
        \hline
     Calibration-free & \multirow{2}{*}{${\times}$} & \multirow{2}{*}{${\times}$} &  \multirow{2}{*}{${\times}$} & \multirow{2}{*}{${\checkmark}$}\\
     within range & {} & {} & {} & {}\\
    \hline
     Sensitivity {[\SI{}{\micro\siemens}]} & {0.01} & {--} & {330e-6} & {40e-6*} \\
   \hline
     Rel. error [\%] & {0.68} & {1}  & {1} & {0.0025*}\\
    \hline
     Power [\SI{}{\micro\watt}] & {--} & {--}  & {620} & {2.3} \\
     \hline
     \end{tabular}
	\footnotesize \raggedright 
	{ * Results from simulations.
    }
 \label{comparison}
\end{table}

\section{Conclusions}
\label{conclusions}

This work has presented VCO-CARE, an ultra-low power AFE for EDA monitoring based on a VCO-based ADC. It enables subject-independent sensing over a 0--20~\SI{}{\micro\siemens} range without runtime calibration. The core 31-phase ring oscillator, implemented in 65nm CMOS and driven by a source follower, consumes an average of 1.2\SI{}{\micro\watt}, contributing to a total post-layout power of 2.3~\SI{}{\micro\watt} including digital logic. Despite a 16\% non-linearity over a 400mV$_\textrm{pp}$ input, its monotonic response and 0.8\SI{}{\micro\volt}$_\textrm{rms}$ noise support reliable amplitude-based feature extraction. The selected configuration ($V_\mathrm{DD} = 0.8$~V, $R_1 = 80$~k$\Omega$) balances sensitivity, range, and safety, achieving 40~pS sensitivity and a relative error below 0. 0025\%, while keeping the current density within safe limits. Although higher $V_\mathrm{DD}$ or lower $R_1$ could extend the range, they would increase power and require a complete recharacterization. Future work will explore adaptive resistor tuning, programmable supplies, and multipath VCOs~\cite{ericSeveralPaths} to improve SNDR and extend its applicability to broader biosignal monitoring. 

\bibliographystyle{IEEEtran}
\bibliography{main}

\begin{thebibliography}{10}
\providecommand{\url}[1]{#1}
\csname url@samestyle\endcsname
\providecommand{\newblock}{\relax}
\providecommand{\bibinfo}[2]{#2}
\providecommand{\BIBentrySTDinterwordspacing}{\spaceskip=0pt\relax}
\providecommand{\BIBentryALTinterwordstretchfactor}{4}
\providecommand{\BIBentryALTinterwordspacing}{\spaceskip=\fontdimen2\font plus
\BIBentryALTinterwordstretchfactor\fontdimen3\font minus \fontdimen4\font\relax}
\providecommand{\BIBforeignlanguage}[2]{{%
\expandafter\ifx\csname l@#1\endcsname\relax
\typeout{** WARNING: IEEEtran.bst: No hyphenation pattern has been}%
\typeout{** loaded for the language `#1'. Using the pattern for}%
\typeout{** the default language instead.}%
\else
\language=\csname l@#1\endcsname
\fi
#2}}
\providecommand{\BIBdecl}{\relax}
\BIBdecl

\bibitem{nagai2019galvanic}
Y.~Nagai \emph{et~al.}, ``{Galvanic Skin Response (GSR)/Electrodermal/Skin Conductance Biofeedback on Epilepsy: A Systematic Review and Meta-analysis},'' \emph{Frontiers in neurology}, vol.~10, p. 377, 2019.

\bibitem{wen2014emotion}
W.~Wen \emph{et~al.}, ``{Emotion recognition based on multi-variant correlation of physiological signals},'' \emph{IEEE Transactions on Affective Computing}, vol.~5, no.~2, pp. 126--140, 2014.

\bibitem{das2016emotion}
P.~Das \emph{et~al.}, ``{Emotion recognition employing ECG and GSR signals as markers of ANS},'' in \emph{2016 Conference on Advances in Signal Processing (CASP)}.\hskip 1em plus 0.5em minus 0.4em\relax IEEE, 2016, pp. 37--42.

\bibitem{dutta2022analysis}
S.~Dutta \emph{et~al.}, ``{An Analysis of Emotion Recognition Based on GSR Signal},'' \emph{ECS Transactions}, vol. 107, no.~1, p. 12535, 2022.

\bibitem{miranda2022bindi}
J.~A. Miranda \emph{et~al.}, ``{Bindi: Affective Internet of Things to Combat Gender-based Violence},'' \emph{IEEE Internet of Things Journal}, 2022.

\bibitem{kappeler2013towards}
C.~Kappeler-Setz \emph{et~al.}, ``{Towards long term monitoring of electrodermal activity in daily life},'' \emph{Personal and ubiquitous computing}, vol.~17, no.~2, pp. 261--271, 2013.

\bibitem{calero2023self}
J.~A.~M. Calero \emph{et~al.}, ``{Self-Adjustable Galvanic Skin Response Sensor for Physiological Monitoring},'' \emph{IEEE Sensors Journal}, vol.~23, no.~3, pp. 3005--3019, 2023.

\bibitem{MITPoh}
M.-Z. Poh \emph{et~al.}, ``{A Wearable Sensor for Unobtrusive, Long-Term Assessment of Electrodermal Activity},'' \emph{IEEE transactions on bio-medical engineering}, vol.~57, pp. 1243--52, 02 2010.

\bibitem{rajeshPPGheartRate}
P.~V. Rajesh \emph{et~al.}, ``{22.4 A 172µW compressive sampling photoplethysmographic readout with embedded direct heart-rate and variability extraction from compressively sampled data},'' in \emph{2016 IEEE International Solid-State Circuits Conference (ISSCC)}, 2016, pp. 386--387.

\bibitem{xing2018wearable}
K.~Xing \emph{et~al.}, ``{A Wearable High-Precision Skin Resistance Acquisition System},'' in \emph{2018 IEEE SmartWorld, Ubiquitous Intelligence \& Computing, Advanced \& Trusted Computing, Scalable Computing \& Communications, Cloud \& Big Data Computing, Internet of People and Smart City Innovation}.\hskip 1em plus 0.5em minus 0.4em\relax IEEE, 2018, pp. 123--127.

\bibitem{kim2020fully}
H.~Kim \emph{et~al.}, ``{Fully Integrated, Stretchable, Wireless Skin-Conformal Bioelectronics for Continuous Stress Monitoring in Daily Life},'' \emph{Advanced Science}, vol.~7, no.~15, p. 2000810, 2020.

\bibitem{banganho2021designAdaptive}
A.~R. Banganho \emph{et~al.}, ``{Design and Evaluation of an Electrodermal Activity Sensor (EDA) with Adaptive Gain},'' \emph{IEEE Sensors Journal}, vol.~21, no.~6, pp. 8639--8649, 2021.

\bibitem{ericVCOinst}
E.~Gutierrez \emph{et~al.}, ``{Why and How VCO-based ADCs can improve instrumentation applications},'' in \emph{2018 25th IEEE International Conference on Electronics, Circuits and Systems (ICECS)}, 2018, pp. 101--104.

\bibitem{gielenLP}
G.~G. Gielen \emph{et~al.}, ``{Time-Encoding Analog-to-Digital Converters: Bridging the Analog Gap to Advanced Digital CMOS-Part 1: Basic Principles},'' \emph{IEEE Solid-State Circuits Magazine}, vol.~12, no.~2, pp. 47--55, 2020.

\bibitem{borgmansFeedforwardRO}
\BIBentryALTinterwordspacing
J.~Borgmans and P.~Rombouts, ``{Enhanced circuit for linear ring VCO-ADCs},'' \emph{Electronics Letters}, vol.~55, no.~10, pp. 583--585, 2019. [Online]. Available: \url{https://ietresearch.onlinelibrary.wiley.com/doi/abs/10.1049/el.2019.0241}
\BIBentrySTDinterwordspacing

\bibitem{quintero}
A.~Quintero \emph{et~al.}, ``{A Coarse-Fine VCO-ADC for MEMS Microphones With Sampling Synchronization by Data Scrambling},'' \emph{IEEE Solid-State Circuits Letters}, vol.~3, pp. 29--32, 2020.

\bibitem{miranda2022wemac}
J.~A. Miranda~Calero, L.~Guti{\'e}rrez-Mart{\'\i}n, E.~Rituerto-Gonz{\'a}lez, E.~Romero-Perales, J.~M. Lanza-Guti{\'e}rrez, C.~Pel{\'a}ez-Moreno, and C.~L{\'o}pez-Ongil, ``Wemac: Women and emotion multi-modal affective computing dataset,'' \emph{Scientific data}, vol.~11, no.~1, p. 1182, 2024.

\bibitem{ericSeveralPaths}
\BIBentryALTinterwordspacing
E.~Gutierrez \emph{et~al.}, ``{N Parallel paths for non-linearity mitigation in ring oscillator-based analog-to-digital conversion},'' \emph{AEU - International Journal of Electronics and Communications}, vol. 170, p. 154778, 2023. [Online]. Available: \url{https://www.sciencedirect.com/science/article/pii/S1434841123002522}
\BIBentrySTDinterwordspacing

\end{thebibliography}

\end{document}